\documentclass[12pt]{article}

\usepackage{latexsym,amsfonts,amsmath,amsthm,amssymb}

\title{A Mathematician's Viewpoint to Bell's theorem:\\
In Memory of Walter Philipp }
\author{Andrei Khrennikov\\
International Center for Mathematical Modeling\\ in Physics,
Engineering and Cognitive science, \\ V\"axj\"o University,
S-35195, Sweden}

\date{}

\begin{document}
\maketitle

\begin{abstract}
In this paper dedicated to the memory of Walter Philipp, we
formalize the rules of classical$\to$ quantum correspondence and
perform a rigorous mathematical analysis of the assumptions in
Bell's NO-GO arguments.
\end{abstract}

\bigskip

\bigskip

I met Walter Philipp on many occasions --- mostly during the
V\"axj\"o conferences and during my visits to the University of
Illinois in Urbana-Champaign --- and I always enjoyed social and
scientific contacts with him. It was impressive that he always
behaved as if he had just moved recently from Vienna (which in
fact he had left as early as the 60th to move to Illinois): he was
a man of great European cultural level, with a deep sense of
humor, and he exhibited them both through uncountable stories
about writers, poets, artists and scientists from Vienna.

Our common  scientific interest was the mathematical formalization
of Bell's arguments \cite{B} which are widely known as Bell's
NO-GO theorem. Contacts with Walter were very attractive for me,
because we  both had the same background: a specialization in
probability theory. I was really happy to find in Walter a fellow
mathematician with whom discussions on Bell's theorem could be
made in the language of mathematically rigorous statements. Walter
and I shared the common viewpoint --- one that I have been trying
to advocate to physicists since the first conference in V\"axj\"o,
on "Foundations of Probability and Physics" \cite{FPP,FPP1,FPP3}
--- that without a rigorous mathematical formalization of the
probabilistic content of Bell's arguments one cannot forcefully
derive the fundamental dilemma that we are often being offered:
that is, {\it either nonlocality or the death of reality}
 \cite{CL,Shim,Wig,AS,HS}.

Our main point was that any mathematical theorem (when formulated
in rigorous mathematical terms) is based on a list of assumptions.
If such a precise list is not provided, then one cannot call it a
mathematical theorem, and should not make any definitive
conclusions. It was pointed out on many occasions, both by Walter
Philipp and his collaborator Karl Hess, see e.g.
\cite{HPL}--\cite{AP}, as well as by Luigi Accardi, see e.g.
\cite{ACC,ACC1}, and myself \cite{AY1}--\cite{AY2}, that without a
presentation of a precise probabilistic model for Bell's
framework, one cannot proceed in a rigorous way.

If one uses the Kolmogorov measure-theoretic model then one should
be aware that there is no reason, even in classical physics, to
assume that statistical data that were obtained in different
experiments should be described by {\it a single Kolmogorov
probability space,} see e.g.  \cite{AY1} for details. Walter
Philipp strongly supported this kind of counterarguments by
finding a purely mathematical investigations in probability which
were devoted to a similar problem, but without any relation to
quantum physics. In particular, Walter found a theorem (proved by
a Soviet mathematician Vorob'ev\cite{VR}) describing the
conditions which are necessary and sufficient for the realization
of a few random variables on a single Kolmogorov space.

We remark that {\it Vorob'ev's theorem} was proved before
the inequality which is nowadays known as Bell's inequality appeared in quantum physics.
Vorob'ev applied his results to game theory. He noticed: "In application to the theory of games
the foregoing means that a combination of players into groups can be found, the mixed
coordinated actions of these groups cannot determine any mixed manner of action for all
the players of the game which is consistent with the actions of these groups. The types of questions,
developed in line with the requirements of the theory of coalition games, can find future
applications also in the theory of information and the theory of random algorithms,"  \cite{VR},
p. 147-148.

Unfortunately,
studies of N. N. Vorob'ev were not supported by mathematical probabilistic establishment. In modern
probability theory people traditionally operate in a single Kolmogorov probability space.
Typically a mathematical paper in probability is started with the sentence:
"Let  ${\cal P}=(\Omega, {\cal F}, {\bf P})$ be a Kolmogorov probability space." Then everything
should happen  in this chosen once and for ever probability space. When I started to develop
so called contextual probability theory, i.e., a theory of probability in that any context
(a complex of physical or biological, or social conditions) determines its own probability space
\cite{CP}, \cite{CP1}, the former students of Kolmogorov, Albert Shiryaev and Alexander
Bulinskii, paid my attention to the fact that Kolmogorov by himself always underlined the role
of a context in determining a probability space of an experiment \cite{K}, \cite{K1}, see also
Gnedenko \cite{G1} and Shiryaev \cite{Sh}.

Walter Philipp and his collaborator Karl Hess demonstrated \cite{HPL}--\cite{AP}
that for some extended models cotaining time as one of hidden
variables the using  of a single Kolmogorov probability space was not justified.

However, in physical literature it was often pointed out that in Bell's framework one need not use
the Kolmogorov measure-theoretic model at all, because {\it
it is possible to operate just with frequencies.}
As was remarked  first by De Baere \cite{Bae} and then by Khrennikov \cite{AY1},  the frequency
derivation of Bell's inequality is also based on mixing of statistical
data from different experiments. Such a procedure was not totally justified.
If we proceed in a rigorous mathematical way by using
the von Mises frequency approach (which was recently
presented on the mathematical level in \cite{AY1}), then we shall immediately see
that it is impossible to obtain Bell's inequality without additional assumptions.
It is amazing that very soon Walter Philipp came independently to the same conclusion.
It seems that everybody who was  educated in probability theory should
see such a problem.

We remark that there can be  obtained generalized
Bell-type inequalities which are not violated by the
experimental statistical data taken from different experiments \cite{AY1}--\cite{AY}.

In this paper I would like to continue the "mathematical line" in analysing Bell's arguments.
I would like to formalize the rules of  classical$\to$quantum correspondence.
And we shall see that it is
not so simple task.
The first step in this direction was done by von Neumann when he  formulated
the first NO-GO statement for existence a prequantum classical statistical model \cite{VN}. J. Bell
continued this activity by starting with heavy critical arguments against von Neumann's
formalization. We proceed in the same way. Our
analysis showed that Bell's formalization was far from to be complete.

I think such a mathematical analysis of Bell's arguments would be the best memorial
in the honor of Walter Philipp.

\section{Derivation of Bell's inequality}

By the Kolmogorov axiomatics \cite{K}, see also \cite{K1},
\cite{G1}, \cite{Sh}  the {\it probability space} is a triple
$$
{\cal P} = (\Omega, {\cal F}, {\bf P} ),
$$
where $\Omega$ is an arbitrary set, ${\cal F}$
is an arbitrary $\sigma$-algebra\footnote{In literature one also uses the terminology
$\sigma$-field, instead of $\sigma$-algebra.} of subsets of $\Omega,$
${\bf P}$ is a $\sigma$-additive measure on
${\cal F}$ which yields  values in the segment $[0,1]$ of the real line
and normalized by the condition ${\bf P}(\Omega)=1.$

\medskip

{\it Random variables} on ${\cal P}$ are by definition measurable functions
\begin{equation}
\label{RV}
\xi: \Omega \to {\bf R}.
\end{equation}
Thus $\xi^{-1}(B) \in {\cal F}$ for every $B \in {\cal B},$
where ${\cal B}$ is the Borel $\sigma$-algebra on the real line.

\medskip

Let  ${\cal P}=(\Omega, {\cal F}, {\bf P})$ be a Kolmogorov
probability space. For any pair of random variables $u(\omega),
v(\omega),$ their covariation is defined  by
$$
<u,v> = \rm{cov}(u,v)= \int_\Omega u(\omega)\;  v(\omega) \; d {\bf
P}(\omega).
$$
We reproduce the proof of Bell's inequality in the measure-theoretic
framework.

\medskip

{\bf Theorem 1.} (Bell inequality for covariations) {\it Let
$\xi_a, \xi_b, \xi_c= \pm 1$ be random variables on ${\cal P}.$
Then Bell's inequality
\begin{equation}
\label{BBB} \vert <\xi_a,\xi_b > - < \xi_c,\xi_b >\vert \leq 1 -
<\xi_a,\xi_c>
\end{equation}
holds.}

{\bf Proof.} Set $\Delta= <\xi_a,\xi_b > - < \xi_c,\xi_b >.$ By
linearity of Lebesgue integral we obtain
\begin{equation}
\label{B1} \Delta = \int_\Omega \xi_a(\omega) \xi_b(\omega) d {\bf
P}(\omega)- \int_\Omega \xi_c(\omega) \xi_b(\omega) d {\bf
P}(\omega)= \int_\Omega [\xi_a(\omega) - \xi_c(\omega)]b(\omega) d
{\bf P}(\omega).
\end{equation}
As
\begin{equation}
\label{LLB}\xi_a(\omega)^2= 1,
\end{equation}
we have:
\begin{equation}
\label{B2} \vert \Delta \vert = \vert \int_\Omega [1 - \xi_a(\omega)
\xi_c(\omega)] \xi_a(\omega) \xi_b(\omega) d {\bf P}(\omega)\vert
\leq \int_\Omega [1 - \xi_a(\omega) \xi_c(\omega)]  d {\bf
P}(\omega).
\end{equation}

\section{Correspondence between classical and quantum models}

It is assumed that there exists a space of hidden variables
$\Omega$ representing states of individual physical systems. There
is fixed a $\sigma$-algebra ${\cal F}$ of subsets  of $\Omega.$ On
this space there are defined {\it classical quantities.} These are
measurable functions $\xi: \Omega \to {\bf R}$ -- random variables
on the measurable space $(\Omega,{\cal F}).$ \footnote{In this
framework classical is equivalent to existence of a {\it functional
representation.} Denote the space of classical quantities by the
symbol $V(\Omega).$ This is some space of real-valued (measurable)
functions on $\Omega.$ The choice of this functional space depends
on a model under consideration. For a system which state is given by
the hidden variable $\omega,$ the value $\xi(\omega)$ of a classical
quantity $\xi$ gives the objective property $\xi$ of this system. We
shall not distinguish a classical (physical) quantity and its
representation by random variable.}

There is also considered a space of physical observables $O$. In the
quantum model they are represented by self-adjoint operators, e.g.,
we can represent  $O$ by ${\cal L}_s({\cal H})$ -- the space of
bounded self-adjoint operators in the Hilbert space of quantum
states ${\cal H}.$ We shall distinguish a physical observable and
its operator-representative by using symbols: $a$ and $\hat{a}.$

The main question is about existence of a correspondence between the
space of random variables $V(\Omega)$ and the space of quantum
observables ${\cal L}_s({\cal H})$ --  about the possibility to
construct a map
\begin{equation} \label{MP} j: V(\Omega)\to {\cal L}_s({\cal
H})
\end{equation}
or a map
\begin{equation}
\label{MPG} i: {\cal L}_s({\cal H}) \to  V(\Omega)
\end{equation}
which have ``natural probabilistic properties'' (in general $j$ is
not a one-to-one map; its existence does not imply existence of $i$
and vice versa).  The main problem is that physics does not tell us
which features such maps should have.  There is a huge place for
mathematical fantasies (presented in the form of various NO-GO
theorems). We now recall the history of this problem.

\section{Von Neumann's formalization}

J. von Neumann was the first who presented a list of possible
features of the classical$\to$quantum map $j,$ see \cite{VN}:

\medskip

VN1). $j$ is one-to-one map.\footnote{Different random variables
from the space $V(\Omega)$ are mapped into different quantum
observables (injectivity) and any quantum observable corresponds to
some random variable belonging $V(\Omega)$ (surjectivity).}

\medskip

VN2). For any Borel function $f: {\bf R} \to {\bf R},$ we have
$j(f(\xi)) = f(j(\xi)), \xi \in V(\Omega).$

\medskip

VN3). $j(\xi_1+ \xi_2+ ...) = j(\xi_1) + j(\xi_2)+...$ for any any
sequence $\xi_k \in V(\Omega).$\footnote{As J. von Neumann remarked:
``the simultaneous measurability of $j(\xi_1), j(\xi_2), ...$ is not
assumed'', see \cite{VN}, p. 314.}

\medskip

Any statistical model contains a space of {\it statistical states.}
In a prequantum statistical model (which we are looking for)
statistical states are represented by probability measures on the
space of hidden variables $\Omega.$ Denote such a space of
probabilities  by $S(\Omega).$ This space is chosen depending on a
classical statistical model under consideration.\footnote{For
example, in classical statistical mechanics $S(\Omega)$ is the space
of {\it all probability measures} on phase space $\Omega={\bf
R}^{2n}.$ In a prequantum classical statistical field theory which
was developed in a series of works \cite{KH1},\cite{KH4} the space
of hidden variables $\Omega$ is infinite-dimensional phase space,
space of classical fields, and the space of statistical states
$S(\Omega)$ consists of Gaussian measures having very small
dispersion.} In the quantum model statistical states are represented
by von Neumann {\it density operators.} This space is denoted by
${\cal D}({\cal H}).$

Roughly speaking J. von Neumann proved that under conditions VN1-VN3
every operation of statistical averaging on $V(\Omega)$ can be
represented as the quantum trace-average on ${\cal L}_s({\cal H})$
corresponding to a quantum  state $\rho \in {\cal D}({\cal H}).$ By
using the language of probability measures we can say that every
probability measure ${\bf P}$ on $\Omega$ can be represented by a
quantum state $\rho$ and vice versa. Thus we have the following
``theorem'' (von Neumann did not proceed in the rigorous
mathematical framework):

\medskip

{\bf Theorem 2.} (von Neumann) {\it Under conditions VN1-VN3 (and
some additional technical conditions) there is well defined map $j:
S(\Omega) \to {\cal D}({\cal H})$ which one-to-one and }
\begin{equation}
\label{AV} \int_\Omega \xi(\omega) d{\bf P}(\omega)= \rm{Tr}\;
\rho\; \hat{a}, \; \mbox{where}\; \rho =j({\bf P}), \hat{a}=j(\xi).
\end{equation}

\medskip

By using Theorem 2 J. von Neumann ``proved'' (he did not formulate
a theorem, but just ansatz) \cite{VN}:

\medskip

{\bf Theorem 3.} (Von Neumann) {\it Let the space of statistical
states $S(\Omega)$ contains probabilities having zero dispersion. A
correspondence map $j$ between a classical statistical model
$$M_{\rm{cl}}=(S(\Omega), V(\Omega))$$ and the quantum statistical
model $$N_{\rm{quant}}=({\cal D}({\cal H}), {\cal L}_s({\cal H}))$$
satisfying the postulates V1--V3 (and some additional technical
conditions \cite{VN}) does not exist.}

\section{Bell's NO-GO theorem}

As was pointed out by many outstanding physicists (e.g., by J. Bell
\cite{B} and L. Ballentine \cite{BL}), some of von Neumann
postulates of classical$\to$ quantum correspondence are nonphysical.
In the opposition to von Neumann, in Bell-type NO-GO theorem
different classical quantities can correspond to the same quantum
observable. It is  not assumed that every self-adjoint operator
corresponds to some classical quantity.  It might be that some
self-adjoint operators have no classical counterpart (or even
physical meaning). The postulate V1 was deleted from the list for
classical $\to$ quantum correspondence. The most doubtful postulate
V3 was also excluded from considerations.  It was not assumed that
the V2 holds.

We consider a family of spin operators:
$$
\hat{\sigma}(\theta)=\cos \theta \hat{\sigma}_z + \sin\theta
\hat{\sigma}_x,$$ where $\hat{\sigma}_x, \hat{\sigma}_z $ are Pauli
matrices, $\theta \in [0, 2\pi).$ These operators act in the two
dimensional state space ${\cal H}={\bf C}^2.$ We also consider spin
operators for pairs of 1/2-spin particles:
$\hat{\sigma}(\theta)\otimes I$ and $I \otimes
\hat{\sigma}(\theta).$ They act in the  four dimensional state space
${\cal H}={\bf C}^2 \otimes {\bf C}^2.$

Bell's list of postulates on classical $\to$ quantum correspondence
can be described as following:

\medskip

B1). The image $j(V(\Omega))$ contains spectral projectors for
operators $\hat{\sigma}(\theta)\otimes I$ and $I \otimes
\hat{\sigma}(\theta)$ for pairs of 1/2-spin particles.

\medskip

B2). For any random variable $\xi \in V(\Omega),$ its range of
values $\xi(\Omega)$ coincides with the spectrum of the operator
$\hat{a}=j(\xi).$

B3). The image $j(S(\Omega))$ contains the singlet spin
state\footnote{This state belongs to the four dimensional state
space ${\cal H}={\bf C}^2 \otimes {\bf C}^2.$ By using the tensor
notations we can write $\psi= \frac{1}{\sqrt{2}}[\vert +
>\otimes\vert -> - \vert - >\otimes\vert + >].$}
$$
\psi= \frac{1}{\sqrt{2}}(\vert + >\vert -> - \vert - >\vert + >)
$$
\medskip

Starting with any classical$\to$quantum mapping $j$ we can
construct a map $i$ from quantum observables to classical random
variables  by setting for $\hat{a}\in j(V(\Omega)),$
$$
i(\hat{a})= \xi_a,
$$
where $\xi_a$ belongs to the set of random variables
$j^{-1}(\hat{a}).$ We also construct a map (denoted by the same
symbol $i)$ from the space of von Neumann density operators into the
space of classical probability measures by choosing a probability
measure ${\bf P}_\rho$ belonging the set $j^{-1}(\rho).$  We
emphasize that such maps
$$
i: {\cal L}_s({\cal H}) \to V(\Omega)
$$
and
$$
i: {\cal D}({\cal H}) \to S(\Omega)
$$
are not uniquely defined! We also point out to a purely mathematical
problem of transition from classical$\to$quantum to
quantum$\to$classical correspondence. Such a possibility is based on
the {\it Axiom of Choice.} For example, for quantum observables we
have the collection of sets $j^{-1}(\hat{a})$ of classical random
variables corresponding to self-adjoint operators. We should choose
from each of these sets one random variable and construct a new set
-- the classical image of quantum observables. The using of this
axiom is not commonly accepted in the mathematical community.

We set
$$\xi_{\theta}= i(\hat{\sigma}(\theta)\otimes I), \; \xi^\prime_{\theta}= i(I\otimes \hat{\sigma}(\theta)).
$$
These are classical pre-images of the spin operators for pairs of
1/2-spin particles.

J. Bell also proposed to use the following postulates:
\medskip

B4). For any quantum state  $\rho$ and commuting operators $\hat{a},
\hat{b},\;$ the quantum and classical correlations coincide:
$$
<\xi_{a}, \xi_{b}>_{{\bf P}_\rho} \equiv \int_\Omega \xi_{a}(\omega)
\xi_{b} (\omega) d {\bf P}_\rho(\omega) = < \hat{a} \; \hat{b}>_\rho
\equiv \rm{Tr} \; \rho   \; \hat{a} \hat{b}.
$$

\medskip

B5). For the singlet state $\psi$ and any $\theta,$ random variables
$\xi_{\theta}$ and $\xi^{\prime}_{\theta}$ are anti-correlated:
\begin{equation} \label{CCM2}
\xi_{\theta}(\omega) = - \xi^\prime_{\theta}(\omega)
\end{equation}
almost everywhere with respect to the probability ${\bf P}_\psi.$

\medskip

{\bf Theorem 4.} (Bell) {\it \it Let $\rm{dim}\; {\cal H} = 4.$
Correspondence maps \begin{equation} \label{CCM3} j: V(\Omega) \to
{\cal L}_s({\cal H}) \; \mbox{and}\; j: S(\Omega) \to {\cal D}({\cal
H})
\end{equation}
satisfying the postulates B1-B5 do not exist.}

{\bf Proof.} We apply Bell's inequality, Theorem 1, to random
variables $\xi_{\theta}= i(\hat{\sigma}(\theta)\otimes I), \;
\xi^\prime_{\theta}= i(I\otimes \hat{\sigma}(\theta)) $ and to a
probability measure ${\bf P}_\psi$ corresponding to the singlet
state $\psi:$
$$
\vert <\xi_{\theta_1},\xi_{\theta_2} >_{{\bf P}_\psi} - <
\xi_{\theta_3},\xi_{\theta_2} >_{{\bf P}_\psi} \vert \leq 1 -
<\xi_{\theta_1}, \xi_{\theta_3}>_{{\bf P}_\psi}.
$$
We remark that the postulate B2 was used here. To prove Bell's
inequality, we took into account that random variables
$\xi_{\theta}(\omega)=\pm 1.$ We now apply the anti-correlation
postulate B5 and rewrite the Bell's inequality:
$$
\vert <\xi_{\theta_1},\xi^\prime_{\theta_2} >_{{\bf P}_\psi} - <
\xi_{\theta_3},\xi^\prime_{\theta_2} >_{{\bf P}_\psi} \vert \leq 1 +
<\xi_{\theta_1}, \xi^\prime_{\theta_3}>_{{\bf P}_\psi}.
$$
Finally, we apply the postulate B4 and write quantum covariations,
instead of classical:
$$
\vert \rm{Tr} \; (\psi\otimes\psi) \;(\hat{\sigma}(\theta_1)\otimes
I) \; (I \otimes \hat{\sigma}(\theta_2)) - \rm{Tr} \;
(\psi\otimes\psi) \;(\hat{\sigma}(\theta_3)\otimes I) \; (I \otimes
\hat{\sigma}(\theta_2))\vert
$$
$$
\leq 1+ \rm{Tr} \; (\psi\otimes\psi)
\;(\hat{\sigma}(\theta_1)\otimes I) \;(I \otimes
\hat{\sigma}(\theta_3)).
$$
It is well known that this inequality is violated for a special choice of angles
$\theta_1, \theta_2, \theta_3.$

\medskip

Our attitude with respect to Bell-type NO-GO theorems is similar to
Bell's attitude with respect to others NO-GO theorems -- von
Neumann, Jauch-Piron and Gleasons theorems, -- \cite{B}, p.4-9. As
well as J. Bell did, we could speculate that some postulates about
the correspondence between classical and quantum models (which were
used in Bell-type NO-GO theorems) were nonphysical. There are many
things which can be questioned in Bell's arguments.

\section{The range of values postulate}

The proofs of Bell and Wigner NO-GO theorems were based on the
postulate B2 on the {\it coincidence of ranges of values} for
classical random variables and quantum observables. Moreover, one
can easily construct examples of classical random variables
reproducing  the EPR-Bohm correlations in the case of violation of
B2, \cite{KV}.

Is the postulate B2 really implied by the physical analysis of the
situation? It seems that not at all! Henry Stapp pointed out
\cite{HS}: ` `The problem, basically, is that to apply quantum
theory, one must divide the fundamentally undefined physical world
into two idealized parts, the observed and observing system, but
{\it the theory gives no adequate description of connection between
these two parts.} The probability function is a function of degrees
of freedom of the microscopic observed system, whereas the
probabilities it defines are probabilities of responses of
macroscopic measuring devices, and these responses are described in
terms of quite different degrees of freedom.'' In such a situation
rejection of the range of values condition is quite natural, since,
as was pointed by Stapp, a classical random variable $\xi$ and its
quantum counterpart $\hat{a}=j(\xi)$ depend on completely different
degrees of freedom. Finally, we remark that a classical model
reproducing quantum probabilistic description, but violating B2, was
recently developed, see \cite{KH1}, \cite{KH4}.

\medskip

{\bf Conclusion.} {\it If the range of values postulates (in the
forms V2, B2) are rejected, then the classical probabilistic
description does not contradict quantum mechanics.}

\section{Contextual viewpoint}

In this section we shall present a very general viewpoint on the
role of  contextuality in Bell-type NO-GO theorems. Bell's original
viewpoint  on contextuality was presented in \cite{B}.
The latter contextuality we can call {\it simultaneous measurement
contextuality} or Bell-contextuality. We reserve the term {\it
contextuality} for our general contextuality -- dependence on the
whole complex of physical conditions for preparation and
measurement. We are aware that commonly in literature
Bell-contextuality is called simply contextuality. However, using
such a terminology is rather misleading, because dependence on the
measurements of other compatible observables is just a very special
case of dependence on the general physical context.

As was rightly pointed out by J. Bell, the only reasonable
explanation of his contextuality  is {\it action at the distance.}
Another possibility is often called {\it ``death of reality''}
 \cite{Shim} -- denying the possibility to assign to
quantum systems objective properties (such as the electron spin or
the photon polarization) -- does not sound natural. The observation
of precise (anti-)correlations for the singlet state evidently
contradicts to the latter explanation.

In contrast to Bell-contextuality, in general contextuality does not
imply action at the distance nor ``death of reality.''

\subsection{Non-injectivity of correspondence}

We first concentrate our considerations on the {\it classical
variables $\to$ quantum observables} correspondence. As we remember,
J. Bell (as well as L. Ballentine) criticized strongly the von
Neumann postulate V1.  Both Bell and Ballentine (as well as many
others) emphasized that there were no physical reasons to suppose
(as von Neumann did) that for a quantum observable $\hat{a}$ its
classical pre-image
$$
j^{-1}(\hat{a})=\{\xi\in V(\Omega): j(\xi)=\hat{a}\}
$$
should contain just one random variable.\footnote{ If one consider the quantum
mechanical description as an {\it approximative description,} cf.
[], then it would be quite reasonable to assume that quantum
mechanics cannot distinguish sharply prequantum physical variables.
A few different classical random variables $\xi, \eta,...$ can be
identified in the quantum model with the same operator $\hat{a}=
j(\xi)=j(\eta)=...$ }

We now consider the {\it classical probabilities $\to$ quantum
states} correspondence. In the same way as for variables and
observables there are no physical reasons to assume injectivity of
the map $j:S(\Omega) \to {\cal D}({\cal H}).$
 By saying that we prepared an ensemble of systems with the
fixed quantum state $\rho$ we could not guarantee that we really
prepared the fixed classical probability distribution. The set
$$j^{-1}(\rho)=\{ {\bf P} \in S(\Omega): j({\bf P}) =\rho\}$$ might
have huge cardinality.

We remark
that the derivations of all Bell-type NO-GO theorems were based on
the possibility to select for any quantum state (at least for the
singlet state) one fixed classical probability measure ${\bf P}_\rho
\in j^{-1}(\rho)$ and for any quantum observable (at least for spin
observables) the fixed  random variable $\xi \in j^{-1}(\hat{a}).$

\subsection{Contextual opposition against Bell's approach to NO-GO
theorems}

The crucial counterargument is that at the experimental level in all
Bell-type inequalities one should use data which is obtained in a
few different runs of measurements (at least three, but in the real
experimental framework four), see \cite{CL}, \cite{Bae},
\cite{AY1}. In the light of the above discussion of the
non-injectivity of classical $\to$ quantum correspondence there are
no physical reasons to assume that we would be able to obtain the
same classical probability distribution  and the same classical
random variables (for example, corresponding to spin
observables).\footnote{Even if we use the same macroscopic
preparation and measurement devices, fluctuations of micro
parameters can induce different physical conditions, \cite{Bae},
\cite{AY1}.} We are not able to guarantee that all runs of
measurements are performed under the same physical conditions.

Let us consider a new random variable $C$ describing a {\it complex
of physical conditions} (context) during a run of measurements.  And
let us try to proceed as J. Bell and his followers did by proving
inequalities for correlations and probabilities. Now classical
probability measures corresponding to a quantum state $\rho$ (in
particular, to the singlet state) depend on runs $C: {\bf
P}_{\rho}\equiv {\bf P}_{\rho,C}$ as well as random variables:
$\xi_{a,C}(\omega), \xi_{b,C}(\omega), \xi_{c,C}(\omega).$ We start
with the correlation inequality. There are three different complexes
of physical conditions $C_1, C_2, C_3$ inducing correlations  which
were considered in  Theorem 1. Here
$$
<\xi_a,\xi_b >(C_1)= \int_{\Omega}  \xi_{a,C_1}(\omega)
\xi_{b,C_1}(\omega) {\bf P}_{\rho,C_1}(\omega),
$$
$$
< \xi_c,\xi_b> (C_2) = \int_{\Omega}  \xi_{c,C_2}(\omega)
\xi_{b,C_2}(\omega) {\bf P}_{\rho,C_2}(\omega).
$$
If $C_1\not= C_2$ we are not able to perform operations with
integrals which we did in Theorem 1. We can not obtain the Bell's
inequality involving the third correlation:
$$
<\xi_a,\xi_c>(C_3)= \int_{\Omega}  \xi_{a,C_3}(\omega)
\xi_{c,C_3}(\omega) {\bf P}_{\rho,C_3}(\omega)
$$
for a context $C_3.$ To derive Bell's inequality, we should assume
that $\label{ASS} C_1= C_2=C_3.$

By using the contextual framework we derived in \cite{KHB},
\cite{AY1} generalizations of the Bell-type inequalities. Such
generalized inequalities do not contradict to predictions of quantum
mechanics. We also mention that a special form of contextuality (so
called {\it non-reproducibility condition}) was also present in
arguments of De Baere \cite{Bae} against Bell's NO-GO theorem, see
also \cite{AY1}. So called efficiency of detectors (or more general
unfair sampling) argument \cite{Gis} can also be considered as
special forms of contextuality -- different contexts produce samples
with different statistical properties.

{\bf Conclusion.} {\it The main value of Bell's arguments was the
great stimulation of experimental technologies for working with
entangled photons.}

\end{document}